\documentstyle[twocolumn,
]{article}
\def\QED{\ \hbox{\ \rlap{$\sqcap$}$\sqcup$}}
\def\join{\mathop{\triangleright\kern-2.2pt\triangleleft}}

\def\QED{\hbox{\rlap{$\sqcap$}$\sqcup$}}

\def\qed{\ifmmode\eqno\QED\else{\parfillskip=0pt \enspace\hfill
  \rlap{$\sqcap$}$\sqcup$ \ifdim\lastskip<\medskipamount
  \removelastskip\penalty55\medskip\fi \parfillskip=0pt
  plus1fil}\fi}
\def\denseformat{
\setlength{\textheight}{9in}
\setlength{\textwidth}{6.9in}
\setlength{\evensidemargin}{-0.2in}
\setlength{\oddsidemargin}{-0.2in}
\setlength{\headsep}{10pt}
\setlength{\topmargin}{-0.3in}
\setlength{\columnsep}{0.375in}
\setlength{\itemsep}{0pt}
\setlength{\parsep}{0pt}
\setlength{\partopsep}{0pt}
\setlength{\topsep}{0pt}
\setlength{\parskip}{0pt}
\setlength{\jot}{0.2pt}
\setlength{\abovedisplayskip}{0.5pt}
\setlength{\abovedisplayshortskip}{0.3pt}
\setlength{\belowdisplayskip}{0.5pt}
\setlength{\belowdisplayshortskip}{0.3pt}
}
\denseformat

\newtheorem{rem} {Remark}            
\newtheorem{notation} {Notation} 
\newtheorem{definition}{Definition}
\newtheorem{thm}{Theorem}

\newtheorem{example}{Example}

\begin{document}
\title{Automatic Generation of Simplified Weakest Preconditions for
Integrity Constraint Verification}
\author{{\large Ahmed A\"\i t-Bouziad }\thanks{\ \ Ahmed A\"\i t-Bouziad: LIAFA,  Universit\'e Paris 6,
4 Place Jussieu, 75252 Paris Cedex 5, France, {\tt
  bouziad@liafa.jussieu.fr}
\hfill\break
\mbox{\ $\;$} ${}^{**}\;$   Ir\`ene Guessarian: LIAFA,  Universit\'e Paris 6,
4 Place Jussieu, 75252 Paris Cedex 5, France, {\it address all
correspondence  to }
{\tt  ig@liafa.jussieu.fr}
\hfill\break
\mbox{\ $\;$} ${}^{***}\;$  Laurent Vieille: Next Century Media, Inc.,
12 Av. des Pr\'es, 78180 Montigny le Bretonneux, France
{\tt lvieille@computer.org}}
 \ \      {\large\ \   Ir\`ene Guessarian ${}^{**}$}
   \and  {\large  Laurent Vieille ${}^{***}$}
         }
\date{\today}
\maketitle

\begin{abstract}

Given a constraint $c$ assumed to hold on a database $B$ and an update
$u$ to be performed on $B$, we address the following question: will
$c$ still hold after $u$ is performed? When $B$ is a relational
database, we define a confluent terminating rewriting system which, starting
from $c$ and $u$, automatically derives a simplified weakest
precondition $wp(c,u)$ such that,  whenever $B$ satisfies $wp(c,u)$,
then the updated database $u(B)$ will satisfy $c$,  and moreover $wp(c,u)$ is simplified in the sense
that its computation depends only upon the instances of $c$ that may
be modified by the update. We then extend the definition of a simplified $wp(c,u)$ to
the case of deductive databases; we prove it using fixpoint induction.

\end{abstract}

\section{Introduction}

We assume a constraint $c$, given by a universal sentence,  on a database $B$ and an update
$u$ to be performed on $B$, and we address the following question: will
$c$ still hold after $u$ is performed? When $B$ is a relational
database, we define a terminating rewriting system which, starting
from $c$ and $u$, automatically derives a simplified weakest
precondition $wp(c,u)$ such that, 1. whenever $B$ satisfies $wp(c,u)$,
then the updated database $u(B)$ will satisfy $c$, 2. $wp(c,u)$ is the
weakest such precondition, and 3. the computation of $wp(c,u)$ depends only upon the instances of $c$ that may
be modified by the update. 
The definition of a weakest precondition $wp(c,u)$ ensuring the
safety of
update $u$ with respect to constraint $c$ extends easily to
 deductive databases with recursive rules and constraints, and even
 updates which can add (delete) recursive rules. When the update is an
 insertion update, we give an algorithm which defines a simplified
 weakest precondition.  We will abbreviate weakest precondition into
 wp.  

\smallskip
A large amount of research work has been devoted to optimizing the
verification of
integrity constraints at  transaction commit. These optimizing efforts
use the fact
that the constraints are verified when the transaction starts, so that
the evaluation
can focus on those constraints which can be violated by the updates and
on the data
relevant to the updates and to the  constraints.  Work in this area
started
for relational databases with techniques to simplify domain-independent
first-order formula \cite{n}.  More recently, techniques based on
propagating updates through the rules of deductive databases have been
developed \cite{bdm, sk, lst}.  These methods are well-understood
by now; they have been tested in prototype implementations
and start appearing in commercial products \cite{v98}.  \cite{vbkl} has shown
that when
general formulae are properly rewritten into rules defining intermediate
predicates, and when update propagation is adequately formalized,  the
simplification approach can be seen as a special case of update
propagation.

However, these methods still involve computation at the end of the
transaction. It is  sometimes claimed that, in practice, this negative
effect on transaction commit time  explains why integrity constraints are
rarely used. While there are many applications  where this is not true, such
a negative effect is probably not acceptable in production  systems
where response time is critical.

This is the reason why another line of research has been proposed.  The
objective is
now to take into account  both the integrity constraints {\em  and the
structure of
the transaction  program }, to try and determine at {\em compile-time }
whether executing the program can violate this constraint or not.
Early work in this line of research include  \cite{gm, cb, ss}. For
more
recent work, see \cite{bs, l,lth,ltw,m}.

To illustrate the differences between the two approaches,
consider the constraint: `forall $x$: $p(x) \rightarrow q(x)$' and the transaction
program
(expressed here in a Prolog-like syntax): `$prog(x) :- insert\ p(x),
insert\ q(x)$.'
Running this program with $x=a$ results in  the insertion of both  $p(a)$ and
$q(a)$.
Optimizing methods will avoid checking the constraint on the whole
databases and
will focus on the data relevant to the updates.  A transaction compiling
approach
will determine at program or constraint  compile-time that,   executing
$prog(x)$
with any parameter will never violate this constraint 
whatever  instance is provided for $x$ in
$prog(x)$,
and no transaction-time activity will occur.

Known theoretical results put a limit on what can be expected from such
an approach
\cite{ahv, bgl}.  Further, not all transaction programs are such that
they can
be proved compliant with the constraints at compile-time. However, (1)
it is natural
programming practice to write transaction programs as safe\footnote
{A transaction program is {\em safe} if it preserves the truth of the constraints.} as possible,
and, (2) if the
compile-time check fails, it is always possible to resort to optimizing
techniques.
Finally, simple examples like the one above indicate that it is worth
attempting
to design {\em effective methods} to prove the compliance of transaction
programs
with integrity constraints.

While predicate calculus appears today as the language of choice to
express integrity
constraints, the choice of an update language is more open. For
instance,  \cite{gm} or
\cite{bs} focus on existing general-purpose programming languages, for
which proving
formal properties is notoriously difficult. In this paper, we follow
\cite{l,lth,ltw}  by
choosing a "pure" (no cut!) logic-programming based language, more
easily
amenable to proofs, in particular against constraints expressed in
predicate calculus.
This choice remains practical for database systems, as they have a
tradition of
providing specific procedural  languages, different from general-purpose
programming languages (e.g.,  stored procedures).

A second parameter is the degree of minimality of the language. While
pure theory
would tend to prefer a clean, mininal language, the design of effective
methods is
often facilitated by the use of additional programming contructs. This
is in
particular true of theorem proving techniques as clearly outlined by W.
Bibel in \cite{blsb}. In this paper, we add an {\tt if -
    then - else} operator
to the language of \cite{ltw}.



In this paper, following the approach of \cite{ltw,l,lth}, we apply
techniques coming from program verification and program
transformations, see for instance \cite{dijkstra, ts, pet,ss}.
Several approaches have been taken along these lines. One can either
generate weakest preconditions, as in \cite{ltw,n,bdm,m} or one can 
generate postconditions, as in \cite{bs}. Once the (pre or
post)conditions are generated, usually the task of verifying them is
left to the user: hints to simplify the precondition are given in
\cite{ltw}, the decidability of checking the
 weakest precondition in relational databases is studied in
\cite{m}, while \cite{n} suggests a method for checking the
precondition on only the relevant part of the database (e.g. the `new'
facts produced by an insertion update).  \cite{bs} define
 post-conditions  $post(u,c)$ and  they
  implement a theorem prover based approach to check the
 safety of updates at compile-time: it consists in proving that
 $post(u,c)\Longrightarrow c$ holds, as this is clearly a sufficient
 condition for ensuring that the update is safe. A different approach to
 the constraint preservation problem has been studied in \cite{bda}:
 it consists in constructing generalized transaction schemes which
 ensure that classes of {\it dynamic} constraints are preserved.

\smallskip\noindent{\bf Contribution of the paper.} It is twofold.

1. In the case of relational databases, we define a terminating rewriting
system which, starting from constraint $c$ and update $u$ {\em
  automatically} derives a
  wp ensuring the safety of the update; assuming that $c$ holds, this
  wp is simplified into a formula which  depends only upon  those
instances of subformulas of $c$ that might be modified by the update.
We prove that our simplified wp is simpler 
than the wp
obtained in \cite{ltw,l,lth} in the following sense: 
 our simplified wp is implied by the wp of \cite{lth},
but the converse does not hold. 

2. For deductive databases allowing
for recursion, we describe an algorithm computing an efficient
simplified wp  in the case of insertion updates.

As soon as recursion is allowed, several problems come up:
1. it is undecidable to check if a
transaction preserves a constraint\cite{ahv}, 
2. the wp is usually not
 expressible in first-order logic \cite{bgl}, and moreover, 
3. if
we want to ensure that both the wp and the constraint are expressed in
the same language \cite{bdm, sk, lst, gsuw, lth},
 we  have to assume a language allowing for both
negation and recursion, which severely limits efficient checking of
the truth of the wp.
  Thus, we can only hope for special cases when
the wp can be shown to hold and/or can be simplified.  

\smallskip

Our update language generalizes the language of \cite{ltw} and of
\cite{lth} by allowing for conditional updates;  it is more
expressive than SQL and the languages of \cite{bdm,lst,n}: e.g. in the
relational case, in \cite{bdm}, only elementary single fact
insert/delete updates are considered, whilst we can insert/delete
subsets defined by first-order formulas as in \cite{bd,ltw,l,lth}. Our
simplified wp are simpler than the wp of \cite{ltw,l,lth}.

The paper is organized as follows: our update language is defined in
section 2, the  terminating rewriting system deriving the simplified
wp for relational databases is described in section 3, heuristics for
treating the deductive case are presented in section 4, and finally
section 5 consists of a short discussion.

\section{Update language}\label{def.sec}

In the present section, we define our update language, which is a mild
generalization of  the update language of \cite{ltw}, and we define the
corresponding weakest preconditions.

\subsection{Definitions}
Let ${\cal U}$ be a countably infinite set of constants.

1.  a {\em database} (DB) $B$ is a tuple $ \langle R_{1},R_{2},...,R_{n}
  \rangle$ where $R_{i}$ is a finite relation  of arity ${k_{i}}$ over
${\cal U}$. The corresponding lower-case letters $r_{1},r_{2},...,r_{n}$ denote the predicate symbols naming
relations $R_{1},R_{2},...,R_{n}$; they are called extensional predicates (or EDBs).

2. an {\em update} $u$ is a mapping from $B= \langle
  R_{1},R_{2},...,R_{n} \rangle$ to $u(B)= \langle
  R_{1}',R_{2}',...,R_{n}' \rangle$ where  $R_{i}$ and $R_{i}'$ have
  the same arity.

3. a {\em constraint}  $c$ is a domain independent 
sentence (a closed
first-order formula which is domain independent, 
 see  \cite{tz,vgt}).

4.  formula $f$ is a {\em precondition} for update $u$ and constraint
$c$ if for every $ B$, if $B
\models f$  then $ u(B) \models c$.

5.  formula $f$ is a {\em weaker} than formula $g$ iff for every database $B$
we have $ B \models g  \Longrightarrow B \models f$.
Formula $f$ is a {\em weakest precondition} for update $u$ and
  constraint $c$, if it is weaker than every precondition for $(u,c)$.

\begin{rem}{\rm  
%
In points 4 and 5 above, $f$ is only a {\rm precondition} for $c$, i.e. $ B
\models f  \Longrightarrow u(B) \models c$ {\em regardless} of whether
$B \models c$.}

\end{rem}
Let $R$ be a  relation in database $B$ and let $\Phi (\overline{x})$ be a
first-order formula with free variables
$\overline{x} = x_{1},x_{2},...,x_{n}$.
\noindent
\begin{itemize}
\vspace{-9pt}
\item  $B' = B[R\rightarrow R']$ denotes the result of substituting
  relation $R'$ for relation $R$ in $B$.
\vspace{-9pt}
\item  $c[r\rightarrow r \cup \phi]$ (resp. $c[r\rightarrow r - \phi]$) denotes the result of substituting
$r(\overline{s}) \lor \phi (\overline{s}) $ 
(resp.  $r(\overline{s}) \land \lnot \phi (\overline{s}) $)
for every occurrence of
$r(\overline{s})$ in $c$.
\vspace{-9pt}\item $c[+r\rightarrow r \cup \phi]$ (resp. $c[-r\rightarrow \lnot r \cup \phi]$) denotes  the result of
  substituting $r(\overline{s}) \lor \phi (\overline{s}) $
 (resp. $\lnot r(\overline{s}) \lor \phi (\overline{s}) $) for every 
positive\footnote{An occurrence of $r(\overline{s})$ is positive if it is
  within the scope of an even number of negations. It is negative otherwise.
} (resp. negative)  occurrence  of $r(\overline{s})$  in $c$.

\end{itemize}

\subsection{The update language}
 Let $\Phi$ be
a first-order formula with free variables $\overline{x}$ and $cond$ a
first-order sentence which are domain independent \cite{tz,vgt}. The instructions of our language are defined as follows.
\begin{definition}\label{def.updtlang}
$I_1$.  {\tt foreach $\overline{x}:\Phi (\overline{x})$ do
  $insert_{R}(\overline{x})$}

$I_2$. {\tt  foreach $\overline{x}:\Phi (\overline{x})$ do
  $delete_{R}(\overline{x})$}

$I_3$.  If  $i_{1}$ and $i_{2}$ are instructions then
  $(i_{1};i_{2})$ is an instruction;

$I_4$. {\tt  if $cond$ then $inst1$ else $inst2$} is an instruction.
 The alternative {else $inst2$} is optional.
\end{definition}
For instance, adding tuple $\overline{a}$ in relation $R$, will be
denoted by {\tt foreach} $\overline{x}:
\overline{x}=\overline{a}$ {\tt do} $insert_{R}(\overline{x})$. In the
sequel, we will abbreviate it by:
$insert_{R}(\overline{a})$.
Formula $\Phi$ is called the qualification of the update. 

The update language of \cite{ltw} consists of instructions $I_1, I_2$ and
$I_3$. Our language is thus  more user-friendly,
because of the possibility of conditional updates defined in $I_4$.
Our language is equivalent to the language considered in \cite{lth}:
complex instructions such as 
\begin{eqnarray}
{\tt  foreach }\ \ \overline{x}:\Phi (\overline{x})\ \ {\tt do }\ \  (i_{1};i_{2})\label{comp.updt}
\end{eqnarray}

\noindent where e.g. for $j=1,2$, $i_{j} = \hbox{\tt foreach }\overline{x_j}:\Phi_j
(\overline{x_j})$ \hbox{\tt do }$insert_{R_j}(\overline{x_j})$ will
be  expressed in our language by  $i_0;i'_1;i'_2\,$,  where {\it TEMP} is a
suitable, initially empty, temporary relation, and
$\overline{y_j}=\overline{x}\setminus \overline{x_j}$ for $j=1,2$:
\begin{eqnarray*}
i_0 &=&\  \hbox{\tt foreach } \overline{x}:\Phi (\overline{x})\ \hbox{\tt  do }insert_{TEMP}(\overline{x})\\
i'_j &=&\   \hbox{\tt foreach
  }\overline{x_j}:\Big(\forall\overline{y_j}\
temp(\overline{x})\wedge\Phi_j (\overline{x_j})\Big)\ \hbox{\tt  do
  }\\
&&\quad insert_{R_j}(\overline{x_j}).
\end{eqnarray*}

\noindent
Our language can be extended to allow for complex instructions such as
\ref{comp.updt}, as well as non sequential or parallel combinations of
updates (such as exchanging the values of two relations). On the other
hand, $I_4$ could also be simulated by several instructions of \cite{ltw}.

We briefly describe the semantics of our update language.
$I_1$ (resp. $I_2$) is executed by first evaluating $\Phi (\overline{x})$ thus
producing the set of instances of $\overline{x}$ satisfying $\Phi
(\overline{x})$ in $B$, and then inserting (resp. deleting) from $ R$
these
 instances of $\overline{x}$. 
  $I_3$ is performed by  executing first $I_{1}$ and then $I_{2}$.
 $I_4$ is performed by evaluating condition $cond$ on $B$, if  $cond$ is true
$inst1$ is executed, otherwise $inst2$  is executed. 
Complex instructions such as {\tt if $cond$ then $(i_{1};i_{2})$ else
  $i_3$} are performed by evaluating first  $cond$ and then, if e.g.
$cond$ is true, executing $i_{1}$ and then $i_{2}$, regardless of
whether  $cond$ holds  after executing  $i_{1}$.

Formally, the update $\lfloor i\rfloor(B)$ associated with instruction
$i$ is defined by:

\smallskip
1. $\lfloor I_{1} \rfloor (B)=B[ R\rightarrow R \cup \{ \overline{x} \vert B\models \Phi (\overline{x})\}]$

2. $\lfloor I_{2} \rfloor (B)=B[ R\rightarrow R - \{ \overline{x} \vert B\models \Phi (\overline{x})\}]$

3. $\lfloor I_{3} \rfloor (B)= \lfloor i_{2} \rfloor(\lfloor i_{1} \rfloor(B))$

4. $\lfloor I_{4} \rfloor (B)= 
\left\lbrace \begin{array}{ccl}
\lfloor inst1 \rfloor (B) & if& B \models cond\\
\lfloor inst2 \rfloor (B) & if& B \models \lnot cond
\end{array}
\right.$
\smallskip

To simplify, $\lfloor i\rfloor(B)$ will be denoted by $i(B)$.

We can easily generalize the transformation of \cite{ltw} to obtain weakest
  preconditions for  our language.
\begin{thm}\label{thm1} The formulas obtained by the weakest
  precondition transformation of \cite{ltw} and defined below in 1--4
  are weakest preconditions for  instructions $I_1-I_4$.

 1. $wp(\hbox{\tt foreach }\overline{x}:\Phi (\overline{x})\ \hbox{\tt  do }insert_{R}(\overline{x}),c) =
 c [r \rightarrow r \cup \Phi]$,

 2. $wp(\hbox{\tt foreach }\overline{x}:\Phi (\overline{x})\ \hbox{\tt  do }delete_{R}(\overline{x}),c) = 
c [r \rightarrow r - \Phi]$,

 3.  $wp((i_{1};i_{2}),c) = wp(i_{1},(wp(i_{2},c))$,

 4. $wp( \hbox{\tt   if }cond\ \hbox{\tt  then }inst1\ \hbox{\tt
    else } inst2,c) = \big(cond\wedge wp(inst1,c)\big)\vee\big(\neg
    cond\wedge wp(inst2,c)\big)$.
\end{thm}

\begin{example} \label{ex1}{\rm 
1. Let $c = \forall \overline{x}\  (r(\overline{x}) \rightarrow
 q(\overline{x}))$
and let $i =\ \hbox{\tt   foreach }  \overline{x}:
p(\overline{x}) \ \hbox{\tt do }  insert_{R}(\overline{x})$, then
\begin{eqnarray*}
wp(i,c)&=&\forall \overline{x} ((r(\overline{x}) \lor p(\overline{x})) \rightarrow q(\overline{x}))\end{eqnarray*}

\noindent 2. Let $c= \forall \overline{x} ( r(\overline{x}) \rightarrow q(\overline{x}))$
and let $i=(i_{1};i_{2})$ where
\begin{eqnarray*}
i_{1} &=& \hbox{\tt  foreach } \overline{x} : s(\overline{x})\
\hbox{\tt   do }delete_{R}(\overline{x})\\
i_{2} &=& \hbox{\tt foreach } \overline{y} : p(\overline{y}) \
\hbox{\tt  do } insert_{R}(\overline{y})\end{eqnarray*}

\noindent then
\begin{eqnarray*}
wp(i,c) &=& wp((i_{1};i_{2}),c)
  =  wp (i_{1},wp(i_{2},c))\\
 & = & wp(i_{1},\forall \overline{x} ((r(\overline{x})\lor p(\overline{x}))\rightarrow
q(\overline{x})))\\
 & = & \forall \overline{x} (\big((r(\overline{x}) \land \lnot s(\overline{x})) \lor p(\overline{x})\big)
 \rightarrow q(\overline{x}))
\end{eqnarray*}
}\end{example}

\begin{rem} \label{sim.rem}
{\rm It is noted in \cite{ltw} that, for constraints which are
 universal formulas in conjunctive normal form, one can take advantage
 of the fact that $c$ holds in $B$ to simplify $wp$ into a $wp'$ such that
 for all $B$ satisfying $c$: $B\models wp' \Leftrightarrow B\models wp$.
Consider for instance the update $i =\ \hbox{\tt   foreach }  \overline{x}:
p(\overline{x}) \ \hbox{\tt do }  insert_{R}(\overline{x})$ of Example
 \ref{ex1} 1, with
 $c = \forall \overline{x}\  (r(\overline{x}) \rightarrow
 q(\overline{x}))$
 then we can simplify $wp(i,c)$ as follows
\begin{eqnarray*}
wp(i,c)&=&\forall \overline{x} ((r(\overline{x}) \lor p(\overline{x})) \rightarrow q(\overline{x}))\\
 & \equiv& \forall \overline{x} (r(\overline{x})\rightarrow q(\overline{x})) \land \forall \overline{x} 
( p(\overline{x}) \rightarrow q(\overline{x}))\\
 &=& c \land \forall \overline{x}( p(\overline{x}) \rightarrow q(\overline{x}))
\end{eqnarray*}
 whence $wp' =   \forall \overline{x}( p(\overline{x}) \rightarrow q(\overline{x}))$.
In the next section, we will apply this simplification in a
systematic way, and implement it via a confluent and terminating rewriting system for weakest preconditions.\QED}\end{rem}

\section{Relational databases}

In the present section, we define a confluent terminating rewriting system which, starting
from $c$ and $u$, automatically derives a simplified weakest
precondition $wp(c,u)$.
The idea underlying the simplification is the same as
 in \cite{ltw,lth}: it consists in  transforming the wp into the form
$wp=c\wedge  wp_1$, and to take advantage of the truth of $c$ in the
 initial database to replace $wp$ by $ wp_1$. However, 

1. it is not clear
 in \cite{ltw,lth} how far this simplification is carried on, and

 2. this simplification is carried on by the user. 

\noindent
On the other hand,
 in our case, the simplification 

1. is iteratively 
 applied until no further  simplification is possible, and 

2. is performed automatically.

\medskip

It is noted in \cite{ltw} that constraints involving existential
quantifiers cannot be simplified: e.g. let $c$ be the constraint
$\exists \overline{x} r(\overline{x})$ and let $u$ be the update
$delete_R (\overline{a})$; then $wp(c,u)= \exists \overline{x}\big(
r(\overline{x})\land \neg(\overline{x}=\overline{a})
\big)$ which cannot be simplified.
Hence, for the simplification to be possible, we will assume the
following restrictions to our language:

1. constraint $c$ and conditions $cond$ are universal sentences,  and 

2. the qualifications $\Phi(\overline{x})$  are 
formulas  without 
quantifications.

\noindent Recall that a clause is a sentence of the form $c= \forall
\overline{x} \;(l_{1} \lor l_{2}\lor ...\lor l_{m})$ where  the $l_i$s are
literals.
Without loss of generality, we can restate our hypotheses as
\begin{enumerate}
\vspace{-9pt}\item  constraint $c$ is a single clause; indeed if $c=\forall \overline{x} \;(D_{1} \land D_{2} \land ... \land D_{n})$ is in
  conjunctive normal form, then we can replace $c$ by the $n$
  constraints $\forall \overline{x} D_{i}$ which are clauses and can
  be studied independently.
\vspace{-9pt}\item  the qualifications $\Phi(\overline{x})$  are  
 conjunctions of literals; we can write
  $\Phi(\overline{x})= \Phi_{1}(\overline{x}) \vee
  \Phi_{2}(\overline{x}) \vee ... \vee \Phi_{n}(\overline{x})$  in
  disjunctive normal form, and we replace the instruction qualified by
  $\Phi(\overline{x})$ by a sequence of similar instructions
  qualified by $\Phi_{1}(\overline{x})$,
  $\Phi_{2}(\overline{x})$,\dots, $\Phi_{n}(\overline{x})$.
\vspace{-9pt}\item  the conditions $cond$ are single clauses: assuming as above that
$cond=\forall \overline{x} (D_{1} \land D_{2} \land ... \land D_{n})$ is in
  conjunctive normal form, we let $cond_i=\forall \overline{x} D_{i}$
  and we replace e.g. {\tt  if $cond$ then $inst$} by {\tt  if $cond_1$
    then  (if $cond_2$ then (\dots (if $cond_n$ then $inst$)\dots))}
\end{enumerate}
Note that in many practical cases, constraints are indeed given by
clauses,  even quite simple clauses involving just 2 or 3
disjuncts, and the restrictions on $cond$ and $\Phi(\overline{x})$ are
also satisfied.

We will use the following notation: let $S$ be a set of clauses and
$r$ a  predicate symbol; $Res_r (S)$ is the set of simplified binary
resolvents of pairs of clauses in $S$ such that
\begin{enumerate}
\vspace{-9pt}\item  each resolvent  is obtained by a unification involving $r$.
\vspace{-9pt}\item  if the resolvent contains a literal of the form 
$\neg(\overline{x}=\overline{a})$ (which we write as
$\overline{x}\not=\overline{a}$), then the resolvent is simplified by
deleting that literal and substituting $\overline{a}$ for
$\overline{x}$ in the resolvent.
\end{enumerate}
\begin{example} \label{ex.resolv} {\rm 
1. Let $S=\{\neg r(x,y)\vee q(y,z)\;,\; r(x,y)\vee
  \neg q(x,y)\}$; then $Res_r (S)=\{q(y,z)\vee \neg q(x,y)\}$.

2. Let $S=\{\neg r(x,y)\vee\neg r(x,z)\vee q(y,z)\;,\; r(x,y)\vee
   \big( (x,y)\not=(a,b)\big)\}$; then the binary resolvents obtained by
  unifying over predicate $r$ all pairs of clauses in $S$ are 
\begin{eqnarray*}\{&\neg r(x,z)\vee
  q(y,z)\vee  \big( (x,y)\not=(a,b)\big)\ ,&
\\  &\neg r(x,y')\vee  q(y',y)\vee  \big(
(x,y)\not=(a,b)\big)&\}
\end{eqnarray*}
they are then simplified into $$Res_r (S)=\{\neg r(a,z)\vee q(b,z)
\;,\;  \neg r(a,y')\vee  q(y',b)\}.\QED$$
}\end{example}

The idea governing our simplification method is as follows:
we define a rewriting system with rules of the form 
\begin{eqnarray*} wp(u,c) &\longrightarrow& \bigwedge wp(u,c_i)\\
 wp(u,c) &\longrightarrow& c'\end{eqnarray*}
and with the property that all formulas in a derivation are logically equivalent.
 Rewriting steps consist in computing the instances of the constraint
 which could be modified by the update. Let us define the following
 abbreviations: we write $insert_{R}\Phi$ for update
$I_1:$\   {\tt foreach $\overline{x}:\Phi (\overline{x})$ do
 $insert_{R}(\overline{x})$}, and we write  $r\vee \neg\phi$ instead of
$\forall\overline{x}\big(R(\overline{x}) \vee \neg\Phi (\overline{x})\big)
 $. Similarly, $delete_{R}\Phi$, (resp. $\neg r\vee \neg\phi$) abbreviates
 $I_2:$\   {\tt foreach $\overline{x}:\Phi (\overline{x})$ do
 $delete_{R}(\overline{x})$}, (resp. $\forall\overline{x}\big(\neg R(\overline{x}) \vee\neg \Phi
 (\overline{x})\big)$).
In updates  $insert_{R}\Phi$ or $delete_{R}\Phi$, we may
assume that $r$ does
  not occur in $\Phi$: indeed any occurrence of $r$ in $\Phi$ is
  preventively renamed before  applying our rewriting rules.

Then our rewriting system consists of the  nine rules given in 
Figure \ref{regles.fig}, where
  $I_1$--$I_4$ are defined in Definition \ref{def.updtlang}, $c$ is a clause
 and $c_1,c_2$ are  universal sentences:


\begin{figure}
$$\begin{array}{lrcl}
{\rm R}_1:\quad& wp(insert_{R}\Phi,c) &\longrightarrow& c[+r\rightarrow  r \cup \phi]\\
&&&\!\!\!\!\!\!\!\!
\hbox{ if } Res_r(c, r \vee\neg \phi)=\emptyset\\
&&&\\
{\rm R}_2:\ & wp(insert_{R}\Phi,c)&\longrightarrow& c[+r\rightarrow  r \cup
\phi]\wedge\\ 
&&&\!\!\!\!\!\!\!\!\!\!\!\!\!\!\!\!\!\!\!\!\!\!\!\!\!\!\!\!\!\!\!\!\!\!\!\!\!\!\!\!\!\!\!\!\!\!{\displaystyle 
\bigwedge_{c_j \in
 Res_r (c,r \vee\neg
 \phi)}}\!\!\!\!\!\!\!\!\!\!\!wp(insert_{R}\Phi,c_j) \ \hbox{ otherwise }\\
{\rm R}_3:\ &  wp(delete_{R}\Phi,c)&\longrightarrow&  c[-r\rightarrow \lnot r
\cup \phi]\\
&&&\!\!\!\!\!\!\!\!\!\!\!
\hbox{ if } Res_r(c,\neg r \vee\neg \phi)=\emptyset\\&&&\\
{\rm R}_4:\ & wp(delete_{R}\Phi,c)&\longrightarrow&  c[-r\rightarrow \lnot r\cup \phi]
\wedge \!\!\!\!\!\\
&&&\!\!\!\!\!\!\!\!\!\!\!\!\!\!\!\!\!\!\!\!\!\!\!\!\!\!\!\!\!\!\!\!\!\!\!\!\!\!\!\!\!\!\!\!{\displaystyle 
\bigwedge_{c_j \in Res_r(c,\neg r \vee\neg \phi)}}\!\!\!\!\!\!\!\!\!\!\!\!
wp(delete_{R}\Phi,c_j)\ \hbox{ otherwise }\\
{\rm R}_5:\ & wp(I_3,c)&\longrightarrow&  wp(i_{1},(wp(i_{2},c))\\
{\rm R}_6:\ & \!\!\!\!\!\!\!\! wp(I_4,c)&\!\!\!\!\!\!\!\!\!\longrightarrow& 
\!\!\!\!\!\!\! \!\!\!\!\big(cond\wedge
wp(inst1,c)\big)\\
&&&\!\!\!\!\!\!\!\vee\big(\neg    cond\wedge wp(inst2,c)\big)\\
{\rm R}_7 :\ & wp(u,\neg c)&\longrightarrow& \neg wp(u, c)\\
{\rm R}_8 :\ & wp(u,c_1\wedge c_2)&\longrightarrow& wp(u,c_1)\wedge wp(u,c_2) \\
{\rm R}_9 :\ & wp(u,c_1\vee c_2)&\longrightarrow& wp(u,c_1)\vee
wp(u,c_2)\end{array}$$\caption{Simplification
rules}\label{regles.fig}\end{figure}

\begin{thm} 1. The rules given in Figure \ref{regles.fig} 
 define a terminating and confluent rewriting system. 

2. The weakest precondition generated by our system is in the form
$wp=swp\wedge c'$, with $c'$ such that $c \Longrightarrow c'$; if $c$
holds, $wp$ can   be simplified into the weakest precondition $swp$
which is weaker than $wp$,
 the weakest precondition defined in Theorem \ref{thm1}.
\end{thm}

\noindent{\it Proof idea:} Rules are repeatedly applied till
saturation, i.e. until an explicit form to which no rule applies is
obtained.

1. The termination is proved by structural induction: each subformula $wp(i,c_j)$
derived from $wp(i,c)$ either is in an explicit form, or has a $c_j$
which is strictly smaller than $c$. Confluence follows from the fact
that each $wp(u,c)$ has a {\em unique} derivation (up to the order in which
  the rewritings are applied).
 
2. 
We prove by induction that our rewriting system generates a  wp in
the form $wp=swp\wedge c'$, with $c'$ such that $c \Longrightarrow c'$
holds. Therefore, taking into account that $c$ holds in $B$, we can
simplify $wp$ into $swp$, and $wp \Longrightarrow swp$ holds. Because
$wp$ is equivalent to the weakest precondition of Theorem \ref{thm1},
and because the weakest
preconditions of \cite{ltw,l,lth} and of Theorem \ref{thm1} are
equivalent, $swp$ is simpler than the weakest
preconditions of \cite{ltw,l,lth} and Theorem \ref{thm1}.
Example \ref{ex3.1} 1 shows that $swp \not\Longrightarrow wp$, i.e. our
$swp$ can be strictly simpler than  $wp$.\QED

\begin{example}\label{ex3.1}{\rm 1.  Let $c = \forall x,y,z\  \big((p(x,y)\land q(y,z)) \longrightarrow
 (p(x,z) \lor q(x,z))\big)$
and let $i =\ \hbox{\tt   foreach }  x,y:(x,y)=(a,a) \ \hbox{\tt do }
insert_{P}(x,y)$, i.e. $\Phi(\overline x)$ is $(x,y)=(a,a)$,
 then the method of \cite{ltw} gives 
\begin{eqnarray*}wp(i,c)&\!=&\!\!\forall x,y,z\ \big(((p(x,y)\lor (x,y)=(a,a))\land q(y,z))\\
&&\ \ \longrightarrow
 ((p(x,z)\lor (x,y)=(a,a)) \lor q(x,z))\big)\end{eqnarray*}
 and our method gives the sequence of rewritings:
\begin{eqnarray}
wp(i,c)&\longrightarrow_{R_2}&\!\! (\lnot q(a,z) \lor p(a,z) \lor q(a,z))\nonumber\\
 &&\ \ \land c[+p\rightarrow  p\cup (x,y)=(a,a) ] \nonumber\\
 & \longrightarrow \ & (\lnot q(a,z) \lor p(a,z) \lor q(a,z))\label{cvrai}\\
& \longrightarrow \ & true \label{wpvrai}\end{eqnarray}

\noindent The  simplification of line \ref{cvrai}
is obtained by taking into account the fact that $c$ holds in $B$,
hence $ c[+p\rightarrow  p\cup (x,y)=(a,a)] $ also holds. Because
$\lnot q(a,z)\lor q(a,z) = true$,  our simplified  weakest
 precondition   equivalent to {\it true} and we obtain line
 \ref{wpvrai}.

\smallskip
2. Consider again Example \ref{ex1} 2. We have the
    sequence of rewritings, where we have underlined the rewritten
    term whenever a choice was possible:
\begin{eqnarray} 
wp((i_{1};i_{2}),c)&\longrightarrow_{R_5}& wp (i_{1},wp(i_{2},c))\nonumber\\
&\longrightarrow_{R_2}&  wp\big(i_{1},c\wedge  wp(i_{2},\neg p\vee q)
\big)\nonumber\\
&\longrightarrow_{R_1}&  wp\big(i_{1},c\wedge (\neg p\vee q)\big)\label{cfsim.rem1}\\
&\longrightarrow_{R_8}& wp(i_{1},c)\wedge {\underline{wp\big(i_{1}, (\neg p\vee q)\big)}}\nonumber\\
&\longrightarrow_{R_3}&wp(i_{1},c)\wedge (\neg p\vee q)\nonumber\\
&\longrightarrow_{R_3}&\big(c[-r\rightarrow \neg r\cup s]\wedge (\neg p\vee q)
\big)\nonumber\\ &\longrightarrow_{\phantom{R_3}}&  (\neg p\vee q)\nonumber
\end{eqnarray}

\noindent Note that line \ref{cfsim.rem1} gives us the simplified form of the
weakest precondition of Example \ref{ex1} 1, (see Remark \ref{sim.rem}). The last simplification
is obtained by taking into account the fact that $c$ holds in $B$,
hence $c[-r\rightarrow \neg r\cup s]$ also holds.

\smallskip
3.  Let $c = \forall x,y,z\  \big(\neg p(x,y)\vee  \neg p(x,z) \vee q(y,z)\big)$
and let $i =\ \hbox{\tt   foreach }  x,y:(x,y)=(a,b) \ \hbox{\tt do }
insert_{P}(x,y)$, then our method gives:
\begin{eqnarray} 
wp(i,c)&\longrightarrow_{R_2}& wp (i,\neg p(a,z)\vee q(b,z)
)\nonumber\\
&&\ \ \land wp(i,\neg p(a,y)\vee  q(y,b))\land c\nonumber\\
&\longrightarrow_{R_2}^*&  wp (i, q(b,b))\land \big(\neg p(a,z)\vee
q(b,z) \big) \nonumber\\
&&\ \ \land\big(\neg p(a,y)\vee  q(y,b)\big)\land c\nonumber\\
&\longrightarrow_{R_1}&q(b,b)\land \big(\neg p(a,z)\vee
q(b,z) \big)\nonumber\\
&&\ \ \land \big(\neg p(a,y)\vee  q(y,b)\big)\land c\label{asimplifier}
\end{eqnarray}
(where $\longrightarrow_{R_2}^*$ means that several
$\longrightarrow_{R_2}$
rewriting steps are performed);
\noindent 
assuming $c$ holds in $B$,  we can simplify \ref{asimplifier} into 
$swp=\forall y,z\ q(b,b)\land \big(\neg p(a,z)\vee
q(b,z) \big)\land \big(\neg p(a,y)\vee  q(y,b)\big)$ which is to be
verified in $B$. Assuming $c$ holds in $B$, the methods of
\cite{n,bdm,bd} yield the postcondition 
$\forall y,z\ \big(\neg p(a,z)\vee
q(b,z) \big)\land \big(\neg p(a,y)\vee  q(y,b)\big)$ which must be
verified in the updated database $i(B)$. \cite{bdm,bd} go one step
further: embedding $B$ in a deductive framework, they simulate the
evaluation of the postcondition via predicates {\tt delta} and {\tt
  new} which are evaluated in $B$ before update $i$ is performed.
}\end{example}

\section{Deductive Databases}

We now extend the definition and verification of a weakest
precondition $wp(u,c)$ 
 to the deductive database setting, where both $c$ and $u$ can be
defined by DATALOG programs. We chose DATALOG because it is the best
understood, most usual and simplest setting for deductive databases.
We will first define our framework, the weakest preconditions, and
then we will give heuristics for 

1. proving that the weakest precondition holds without actually evaluating it

2. computing simplified weakest preconditions.

\medskip\noindent Recall that on a language consisting of the EDBs $r_1,\ldots,r_k$
and new predicate symbols  $q_1,\ldots,q_l$ 
--called  intensional predicates (or IDBs)--,
a  DATALOG  program  $P$  is a finite set of function-free Horn
clauses,  called rules, of the form:
$$q(y_1,...,y_n)\longleftarrow
q_1(y_{1,1},...,y_{1,n_1}) \;,\;\ldots  \;,\;q_p(y_{p,1},...,y_{p,n_p} )$$ 
where the
$y_i$s and the $y_{i,j}$s  are either variables or 
constants, $q$  is an intensional predicate in $\{q_1,\ldots,q_l\}$, the
$q_i$s  are either  intensional predicates or extensional predicates.

\medskip\noindent
In our framework, both updates and constraints range over deductive
queries, possibly involving recursion. Formally, updates and
constraints are defined as in Section \ref{def.sec}, but in addition:
\begin{itemize}
\vspace{-9pt}\item  the qualifications
  $\Phi(x)$ in both $insert$ and $delete$ statements  are  
 conjunctions of literals which may contain atoms defined by recursive DATALOG programs,
\vspace{-9pt}
\item  similarly,  constraints $c$ and conditions $cond$ in {\tt if -
    then - else} statements  may contain atoms defined by recursive
    DATALOG programs: constraints and conditions are general clauses,
    but the atoms occurring in them are defined by Horn clauses.
\end{itemize}
\noindent The definition of the weakest preconditions extends easily: we follow
here the approach of \cite{l}. 
\begin{notation}{\rm 1. Let $Q$ and $R$ be 
  relations appearing in a DATALOG program $P$, and   respectively 
corresponding to the  predicate symbols $q$ and $r$: 
$q$ is said to {\em depend directly} on $r$ if  $q=r$ or if $r$ appears in
the body of a rule defining $q$; $q$ is said to {\em  depend}  on $r$ if $q$
depends directly on $r$, or $q$ depends directly on  $q'$ and  $q'$ 
depends  on $r$. 

2. Let $P'_r$ be the program obtained by adding to $P$
  new IDB symbols $q'$ for each predicate
$q$ depending  on $r$, and
 new rules;  for each rule $\rho$ of $P$ defining an IDB
$q$ depending  on $r$,  a new rule $\rho'$ defining $q'$ is added: $\rho'$ is
obtained from $\rho$ by substituting  $s'$ for each occurrence of a
  symbol $s$ depending on $r$ (hence $r'$ is substituted 
for each occurrence of $r$). If $r$
  is an EDB predicate, we add a new IDB symbol $r'$, but  
there is no rule defining $r'$ yet: 
the rules defining $r'$
depend on the update and will be given later.

3. $c[r\rightarrow r']$ denotes $c$ where all the predicates depending on
$r$ (including $r$ when $r$ is an EDB) are replaced by the corresponding primed predicates.\QED}
\end{notation}

We will assume the following hypotheses:

\begin{itemize}
\vspace{-9pt}
\item[$H_1$:]  the qualifications $\Phi(\overline{x})$  are  
 conjunctions of literals, and  all the rules defining the IDBs in
 constraint $c$, qualifications $\Phi(\overline{x})$ and conditions
 $cond$ are given in program $P$. 
\vspace{-9pt}
\item[$H_2$:]  in statements of the form
{\tt foreach $\overline{x}:\Phi (\overline{x})$ do
  $insert_{R}(\overline{x})$},
or 
{\tt foreach $\overline{x}:\Phi (\overline{x})$ do $delete_{R}(\overline{x})$}, none of the literals in $\Phi
 (\overline{x})$ depends on $r$.
\end{itemize}
\begin{thm}\label{wp.ded.thm} Assume constraint $c$ and instruction $i$
 satisfy $H_1$ and  $H_2$, then the formulas defined below are
 weakest preconditions for $c$ and $i$.

 1. $wp(\hbox{\tt foreach }\overline{x}:\Phi (\overline{x})\ \hbox{\tt  do }insert_{R}(\overline{x}),c) =
 c [r \rightarrow r']$
where the program defining the IDBs  is $P'_{insert_{R}\Phi}=P'_r\cup
\{r'(\overline{x})\longleftarrow r(\overline{x})\;,\; r'(\overline{x})\longleftarrow \Phi (\overline{x})\}$,

 2. $wp(\hbox{\tt foreach }\overline{x}:\Phi (\overline{x})\ \hbox{\tt  do }delete_{R}(\overline{x}),c) = 
c [r \rightarrow r']$
where the program defining the IDBs is $P'_{delete_{R}\Phi}=P'_r\cup
\{r'(\overline{x})\longleftarrow r(\overline{x})\land \neg
t(\overline{x})\;,\; t(\overline{x})\longleftarrow \Phi
(\overline{x})\}$, with $t$ a new IDB predicate,

 3.  $wp((i_{1};i_{2}),c) = wp(i_{1},(wp(i_{2},c))$,

 4.  $wp( \hbox{\tt   if }cond\ \hbox{\tt  then }inst1\ \hbox{\tt
    else } inst2,c) = \big(cond\wedge wp(inst1,c)\big)\vee\big(\neg
    cond\wedge wp(inst2,c)\big)$.
\end{thm}
Theorem \ref{wp.ded.thm} calls for some remarks.

1. When none of $r$, $c$ or $\Phi$ is recursive, we obtain again the
weakest preconditions  of Theorem \ref{thm1}.

2. Our definition of insertions is quite liberal, allowing us to add
new rules, which is not permitted in \cite{l}. Similarly, deletions can suppress tuples, sets of tuples
or even rules.

3. Deletions and/or qualifications $\Phi$ containing negations 
force us out of the DATALOG framework, because the
weakest precondition of $\hbox{\tt foreach }\overline{x}:\Phi
(\overline{x})\ \hbox{\tt  do }delete_{R}(\overline{x})$ and
constraint $c$ is defined by the DATALOG${}^\neg$ program $P'_r\cup
\{r'(\overline{x})\longleftarrow r(\overline{x})\land \neg
t(\overline{x})\;,\; t(\overline{x})\longleftarrow \Phi
(\overline{x})\}$; this was already noted in \cite{gsuw}. In
\cite{l,lst}  a stratified DATALOG${}^\neg$ 
 framework is assumed: this ensures that both the constraint and the wp are
expressible in the same  framework.

4. In what follows, we will consider only insertions and positive 
qualifications in order to be
able to express both constraints and their wps in DATALOG. The $wp$s
defined in Theorem \ref{wp.ded.thm} are correct without this restriction, but
they are defined by stratified DATALOG${}^\neg$ programs, and not by
Horn clauses.

\begin{example}\label{ex4}{\rm  Let $c$ be the constraint $\forall x,y\allowbreak\ \lnot
  tc(x,y)\vee \allowbreak i(x,y)$, where $i$ does not depend on $tc$, and consider
  the update $\hbox{\tt foreach }{x,y}:path (x,y)\
  \hbox{\tt  do }\allowbreak insert_{TC}(x,y)$, where all the predicates
  are defined by program $P$:\penalty -5000
$$P \ \left\lbrace \begin{array}{lcl}
tc(x,y) &\longleftarrow& arc(x,y)\\
tc(x,y)&\longleftarrow& arc(x,z)\;,\; tc(z,y)\\
path(x,y) &\longleftarrow& edge(x,y)\\
path(x,y)&\longleftarrow& edge(x,z)\;,\; path(z,y)\\
i(x,y)&\longleftarrow& body(x,y)
\end{array}\right.$$
Then $P'_{tc}$ is $P$ together with a new predicate $tc'$ and the rules
\ref{reg'1} and \ref{reg'2}.
\begin{eqnarray}
tc'(x,y) &\longleftarrow& arc(x,y)\label{reg'1}\\
tc'(x,y)&\longleftarrow& arc(x,z)\;,\; tc'(z,y)\label{reg'2}
\end{eqnarray} 
 and $P'_{insert_{TC}path}$
is $P'_{tc}$ together with the  rules \ref{reg''1} and \ref{reg''2}.
\begin{eqnarray}
tc'(x,y) &\longleftarrow& tc(x,y)\label{reg''1}\\
tc'(x,y)&\longleftarrow& path(x,y)\label{reg''2}
\end{eqnarray} 
Finally, $wp(u,c)=\forall x,y\ \lnot
  tc'(x,y)\vee i(x,y)$, where $TC'$ is defined by
  $P'_{insert_{TC} path}=P\cup \{\ref{reg'1} ,\ref{reg'2},\ref{reg''1} ,\ref{reg''2}\}.$
}\end{example}

\smallskip 

We now turn our attention towards the goal of
 proving that the weakest precondition holds without actually evaluating it.
One method is to show that 
\begin{eqnarray} 
c\Longrightarrow wp(u,c)\label{c.implique.wp}
\end{eqnarray} 
The problem is that implication \ref{c.implique.wp} is undecidable
except in some special cases: e.g., if both $c$ and $wp(u,c)$ are unions of conjunctive queries,
and at least one of them is not recursive \cite{c,cv};  some
 special classes of formulas for which \ref{c.implique.wp} is
decidable are studied in \cite{m}. So we can only hope for heuristics to
find sufficient conditions ensuring that implication
\ref{c.implique.wp} will hold.
The idea, coming from Dijkstra's loop invariants \cite{dijkstra2},
consists in proving \ref{c.implique.wp} by recursion induction,
without actually computing $wp(u,c)$. We illustrate this idea on an
example.
\begin{example} {\rm  Let $c$ be the constraint $\forall x,y\ \lnot
  tc(x,y)\vee i(x,y)$, where $I$ and $TC$ are defined by $P$:
\begin{eqnarray}
tc(x,y) &\longleftarrow & arc(x,y)\label{eq1}\\
tc(x,y)  &\longleftarrow & edge(x,y)\label{eq2}\\
tc(x,y) &\longleftarrow & arc(x,z)\;,\; tc(z,y)\label{eq3}\\
i(x,y) &\longleftarrow & \! i_1(x,z_1),edge(z_1,z_2), i_1(z_2,y)\label{eq7}\\
i_1(x,y)  &\longleftarrow & arc(x,z)\;,\;i_1(z,y)\label{eq4}\\
i_1(x,y)  &\longleftarrow & edge(x,z)\;,\;i_1(z,y)\label{eq5}\\
i_1(x,x) &\longleftarrow & \label{eq6}
\end{eqnarray}
and consider
  the update $u=\hbox{\tt foreach }{x,y}:edge (x,y)\
  \hbox{\tt  do }insert_{Arc}(x,y)$. Then $wp(u,c)=\forall x,y\ \lnot
  tc'(x,y)\vee i(x,y)$, where $I$ and $TC'$ are defined by
  $P'_{insert_{Arc}edge}$, consisting of $P$ together with the rules
  (because $I'=I$ here):


\begin{eqnarray*}
arc'(x,y) &\longleftarrow & arc(x,y)\\
arc'(x,y)  &\longleftarrow & edge(x,y)\\
tc'(x,y) &\longleftarrow & arc'(x,y)\\
tc'(x,y) &\longleftarrow & arc'(x,z)\;,\; tc'(z,y)
\end{eqnarray*}
We prove that $c\Longrightarrow wp(u,c)$ by induction. To this end,
let $C[R]= (R\subset I)$; we note  that $c$ holds iff $C[TC]$
holds, and  $wp(u,c)$ holds iff $C[TC']$
holds. Let $\join$ denote the
composition\footnote{$\join$ performs an equijoin on the second
  attribute of the first relation and the  first
  attribute of the second relation, followed by a projection on the
  first and third attributes.} of binary relations. 
 It thus suffices to prove that, for
any $R$, $C[R]\Longrightarrow C[Arc'\cup Arc'\join R]$ to conclude, by
fixpoint induction, that $C[TC']$ holds. 

We now show that, if $c$
holds, then $C[R]\Longrightarrow C[Arc'\cup Arc'\join R]$ holds. 
 Assume that $C[R]$ holds. Because $c$ holds, $C[TC]$ holds. Note that
$C[Arc'\cup Arc'\join R]$ reduces to the conjunction of $C[Arc]$,
$C[Arc\join R]$, $C[Edge]$, $C[Edge\join R]$; each of the conjuncts is
easy to verify: for instance $C[Arc]$ holds because $Arc\subset TC$ by
rule \ref{eq1} and because $C[TC]$ holds; $C[Arc\join R]$ holds
because $C[R]$ holds, and because of rules \ref{eq7}, \ref{eq4}, \ref{eq7}, and
similarly for $C[Edge]$ and $C[Edge\join R]$.
\QED}\end{example}

\medskip

We now study the computation of  simplified weakest preconditions, in
the case of insertion updates.
The basic idea is quite simple and comes from the semi-naive query
evaluation  method in DATALOG (see \cite{ahv}). We design a DATALOG program computing
all new facts deduced from the insertion update (and preferably only
new facts) and we verify the  constraint on the new facts computed by
that  program. The
 method of \cite{bd} is based on a similar idea. We will sketch this
 method on an example, simple, but useful, where the constraint is 
$c= \ \lnot\exists
x\  tc(x,x)$ where $TC$ is a transitive closure.
Such a constraint is used, for instance, to check that a set of
DATALOG clauses defines a  non recursive
program by verifying that the precedence graph of the IDBs occurring
in the program has no cycle. We want to check that adding a new clause
does not create recursions, i.e. cycles. Adding a new clause
corresponds to an insertion update.

\begin{example}\label{ex6} {\rm  Consider
  the update $insert_{Arc}{(d,b)}$, and let $c$ be the constraint 
$\neg\exists x\  tc(x,x)$,  where $TC$  is defined by program $P$:
$$P \ \left\lbrace \begin{array}{lcl}
tc(x,y) &\longleftarrow& arc(x,y)\\
tc(x,y)&\longleftarrow& arc(x,z)\;,\; tc(z,y)\end{array}\right.$$
We assume that constraint $c$ is verified by database $B$.  
Let $\Delta_{tc}(x,y)$ be
the potentially new facts which will be inserted in $TC$ as a consequence of the
update. The IDB predicate $\delta_{tc}$ corresponding to $\Delta_{tc}$ is
defined by the DATALOG program:
$$P' \ \left\lbrace
\begin{array}{lll}
tc(x,y) &\longleftarrow& arc(x,y)\\
tc(x,y)&\longleftarrow& arc(x,z)\;,\; tc(z,y) \\
\delta_{tc}(d,b)& \longleftarrow & \label{cni1}\\
\delta_{tc}(d,y)& \longleftarrow & tc(b,y) \label{cni2}\\
\delta_{tc}(x,y)&\longleftarrow & arc(x,z)\;,\; \delta_{tc}(z,y) \label{cni3}
\end{array}\right.$$The weakest precondition
$wp(c,insert_{Arc}{(d,b)})$ then is $\neg\exists x\  \delta_{tc}(x,x)$,
which can be evaluated by SLD-AL resolution \cite{v}.  The weakest precondition of \cite{l} and of Theorem
\ref{wp.ded.thm} would be in the present case $\neg\exists x\ 
tc'(x,x)$ where $TC'$ is defined by the program
$P'_{insert_{Arc}{(d,b)}}$:
$$P'_{insert_{Arc}{(d,b)}} \ \left\lbrace
\begin{array}{lll}
tc(x,y) &\longleftarrow& arc(x,y)\\
tc(x,y)&\longleftarrow& arc(x,z)\;,\; tc(z,y) \\
tc'(x,y) &\longleftarrow& arc'(x,y)\\
tc'(x,y)&\longleftarrow& arc'(x,z)\;,\; tc'(z,y) \\
arc'(x,y) &\longleftarrow& arc(x,y)\\
arc'(d,b) &\longleftarrow& 
\end{array}\right.$$}\end{example}
The program $P'$ of Example \ref{ex6} can
be obtained by an algorithm: the idea is to compute the rules defining
new facts by resolution with the inserted atoms (similar to the idea
of `refutation with update as top clause' of \cite{sk}). A saturation
method \cite{ahv,bdm,sk,lst} is used to generate the new rules,
i.e. we add rules until nothing new can be added.
To simplify the notations, we give the algorithm in the case when the
update is of the form $insert_R(\overline {d_j})$ for $j=1,\ldots,k$, with $r$ an
EDB predicate, and the
constraint 
is of the form $\neg\exists \overline x \ t( \overline x)$ 
with $t$ an IDB, possibly depending on $r$, defined by a linear DATALOG program.

\medskip
{\bf Algorithm.} {\em Inputs: } update $u= insert_R(\overline {d_j})$ for $j=1,\ldots,k$, 
constraint $c=\neg\exists \overline x \  t(\overline x)$, and a linear
DATALOG program $P$ defining relation
$T$.

\noindent {\em Outputs: } A simplified $wp(u,c)$,
and a DATALOG program $P'$ defining $wp(u,c)$.

{\em Step 1: } For each IDB $q$ of $P$ depending on $r$, let
$\delta_q$ be a new IDB; let $\Pi$ be the set of rules defined as
follows: for each rule $q\longleftarrow body$ of $P$ whose head $q$
depends on $r$ add in $\Pi$ a rule  $\delta_q\longleftarrow body$.
If $\Pi$ is empty, then update $u$ is safe; STOP.

{\em Step 2: } Let $P_1=\{Res(\rho,r(\overline {d_j})) \ /\ \rho\in\Pi,
j=1,\ldots,k\}$ be the set of resolvents of rules of $\Pi$ with
updated atoms.
If $P_1$ is empty, then update $u$ is safe; STOP.

\noindent Otherwise, generate new rules as follows:

$i:=1$ WHILE $P_i\not= \emptyset$ DO $P_{i+1}=\{Res(\rho,r(\overline {d_j}))
\ /\ \rho\in P_i, j=1,\ldots,k\}$\ ;  \ $i:=i+1$ ENDDO

Let $P'=\cup P_i$.

{\em Step 3: } Let $\Sigma_1$ be the set of rules of $\Pi$ which
contain an IDB $q$ depending on $r$ in their body.
If $\Sigma_1$ is empty, then $wp(u,c)= \neg\exists \overline x \ 
\delta_t(\overline x)$,
and program $P\cup P'$ defines $\Delta_t$; STOP.

\noindent Otherwise, generate new rules as follows: let $P''_1$ be
obtained from $\Sigma_1$ by substituting $\delta_q$ for $q$ in the
bodies of the rules of $\Sigma_1$; only new predicates $\delta_q$
appear in rules of $P''_1$.

$i:=1$ WHILE $P''_i\not= \emptyset$ DO $P''_{i+1}=\{Res(\rho,r(\overline {d_j}))
\ /\ \rho\in P''_i, j=1,\ldots,k\}$\ ;  \ $i:=i+1$ ENDDO

Let $P''=\cup P''_i$.

$wp(u,c)= \neg\exists \overline x \  \delta_t(\overline x)$,
and program $P\cup P'\cup P''$ defines $\Delta_t$; STOP.\QED

\smallskip
The WHILE loops in steps 2 and 3 terminate, because at each iteration
step the number of atoms involving $r$ decreases in the rules.
This algorithm can generate  rules which might be useless in some cases: e.g., in
Example \ref{ex6}, the useless rule $\delta_{tc}(d,y) \longleftarrow
 \delta_{tc}(b,y)$ would be generated.
This algorithm can be generalized to non linear DATALOG programs, and
to more general $insert$ instructions.

\smallskip
When the insertion is defined by a (possibly recursive) qualification,
we can similarly compute the potentially 
new facts to be inserted  as a consequence of the
update. Consider the program $P$ and the update 
$u=\hbox{\tt foreach }{x,y}:path (x,y)\
  \hbox{\tt  do }insert_{TC}(x,y)$ defined in
Example \ref{ex4}, and let $c$ be constraint $\neg\exists x\  tc(x,x)$.
The potentially new\footnote{ Some new facts could be already present in the old
  database or could be generated twice; this is unavoidable, unless we are
  willing to perform a semantical analysis of the program which can 
 be expensive.} facts $\Delta_{tc}(x,y)$ which will be inserted in $TC$ are
defined by the DATALOG program $P'$:
$$P' \ \left\lbrace
\begin{array}{lll}
P && \\
\delta_{tc}(x,y)& \leftarrow & edge(x,y)\\
\delta_{tc}(x,y)&\leftarrow & edge(x,z)\;,\; tc(z,y)\\
\delta_{tc}(x,y)&\leftarrow & edge(x,z)\;,\; \delta_{tc}(z,y)\\
\delta_{tc}(x,y)&\leftarrow & arc(x,z)\;,\; \delta_{tc}(z,y)
\end{array}\right.$$
The weakest precondition
$wp(c,insert_{Arc}path)$ is again $\neg\exists x\  \delta_{tc}(x,x)$.

The method of  \cite{l} and of Theorem
\ref{wp.ded.thm} would now give the weakest precondition  $\lnot\exists x\ 
tc'(x,x)$ where $TC'$ is defined by the program
$P'_{insert_{Arc}edge}$, namely:
$$P' \ \left\lbrace
\begin{array}{lll}

P && \\
tc'(x,y) &\longleftarrow& arc'(x,y)\\
tc'(x,y)&\longleftarrow& arc'(x,z)\;,\; tc'(z,y) \\
arc'(x,y) &\longleftarrow& arc(x,y)\\
arc'(x,y) &\longleftarrow& path(x,y)
\end{array}\right.$$

\section{Conclusion and discussion}

In the relational case, we devised a systematic method for computing a simplified weakest precondition for a
general database update transaction $u$ and a constraint $c$. This yields an
efficient way of ensuring that the update maintains the truth of the
 constraints. In the deductive case, we studied two methods: the first
 one consists in proving by fixpoint induction that $c\Longrightarrow wp(u,c)$
holds without evaluating $wp(u,c)$; the second one consists in
defining, for insertion updates and constraints $c$ of the form
$\neg\exists \overline x\  q(\overline x)$, a constraint $c'$ simpler than
$wp(u,c)$ and such that $c\Longrightarrow \Big(c'\Longleftrightarrow
wp(u,c)\Big)$; this is a first step towards one of the goals stated in
the conclusion of \cite{bgl}.

The idea of our method is to preventively check only relevant parts of
  the precondition which are generated using saturation methods. We preventively check a
  weakest precondition {\it before} performing the update, and perform
  the update only when the weakest precondition ensures us that it
  will be safe (see also \cite{bdm,ltw,l,lth}); complex
  updates are also considered a whole, rather than separately, thus
  generating simpler weakest preconditions.
 Following the method initiated in
  \cite{n}, we check only the relevant part of the weakest precondition (i.e. those
  facts potentially affected by the update); to this end, we
  preventively simplify the weakest precondition by using a resolution method: we
  separate in the weakest precondition the facts which reduce to $c$ (assumed to hold)
  from the `new' facts which have to be checked (see also \cite{bdm,n,sk}).

Our update language is more expressive than the ones considered in
 \cite{bs,bdm,lst,n} in that we allow for 1. more complex updates, inserting or
 deleting sets defined by a qualification which is a universal
 formula, 2. conditional updates,  3. complex transactions, and
 4.  recursively defined updates and constraints. The
 language of \cite{bs} has an additional statement {\tt forone  $x$
 {where} $cond$ {do} $inst$} which can be simulated in our language;
 in addition it is object-oriented, as are the languages of \cite{l,lth}.
Our update language is in some respects more user-friendly than the one
 considered in  \cite{l,lth} (because we allow for conditional
 updates, and, in the deductive case, we allow for insertions or
 deletions of rules); in some respects it is less expressive (because our
 qualifications are universal formulas instead of arbitrary first
 order formulas in \cite{l,lth,m}); our system has been extended to
 allow for some existential quatifiers and in practice, our
 qualifications suffice to model usual update languages.   This
 slight loss in expressivity enables us to explicitly and effectively
 give an
 automatic procedure for generating a simplified weakest precondition, implemented via a
 terminating term rewriting system. \cite{ltw,lth} give only {\it
 sufficient} conditions under which simplifications are possible, and
 state the existence of a simplified weakest precondition, without giving an algorithm
 to compute it. 

The parallel time complexity for computing $wp(u,c)$ is linear in the
total size of the formulas involved (constraint, qualification,
etc.). The maximum size of $wp(u,c)$ is also linear in the
 size of the formulas involved, except for the case of an update of
 the form  $insert_R\Phi$ paired with a constraint $c$ containing
 $k>1$ occurrences of $\neg r$, when a blow-up exponential in $k$ may
 occur in the size of $wp(u,c)$ (and similarly for $delete_R\Phi$ with
  $k>1$ occurrences of $ r$ in $c$).


Because our weakest precondition is defined independently of whether
$c$ holds in $B$, and is then simplified by taking into account
whether $c$ holds in $B$, our approach can be extended to handle
changes in the integrity constraints. 
Further steps would be:
\begin{enumerate} \item to apply semantic query optimization
techniques for recursive programs \cite{cgm,jm} to simplify even more
our  simplified weakest preconditions; 
\item to incorporate in our language 
 complex updates (e.g. modifications, exchanges);
\item to generalize our algorithms to allow for constraints which are not
given by clauses. 
\end{enumerate}

\end{document}